\documentstyle[aas2pp4]{article}
\input psfig.tex
\def \m{\ifmmode M_\odot\else M$_\odot$\fi}
\def \r{\ifmmode R_\odot\else R$_\odot$\fi}

\def \gta {\mathrel{\vcenter
     {\hbox{$>$}\nointerlineskip\hbox{$\sim$}}}}

\def\kms{km~s$^{-1}$}

\def\beq{\begin{equation}}
\def\eeq{\end{equation}}
\def\ref{\reference}
\def\gr{$\gamma$-ray }
\def\grb{$\gamma$-ray burst }
\def\grbs{$\gamma$-ray bursts }

\begin{document}
\title{Aspherical Explosion Models for SN 1998bw/GRB 980425} 
\author{ Peter H\"{o}flich, J. Craig Wheeler,  Lifan Wang}
\affil{Department of Astronomy, University of Texas, Austin, TX 78712, USA}   
\affil{E-Mail: pah@alla.as.utexas.edu, wheel@alla.as.utexas.edu, lifan@tao.as.utexas.edu}

\begin{abstract}
 The recent discovery of the unusual supernova SN~1998bw and 
its apparent correlation with the
\grb\ GRB~980425 has raised new issues concerning both \grbs\ 
and supernovae. Although the spectra of SN~1998bw
resemble those of Type~Ic
supernovae (SN~Ic), there are distinct differences.  At early
times the expansion velocities 
inferred by the Doppler shift of (unidentified) absorption
features were very high, and SN~1998bw was unusually bright and 
red at maximum light (Gamala et al. 1998).  
These distinctions make SN~1998bw a candidate 
for a ``hypernova" with explosion energies exceeding normal
supernovae by a factor in excess of 10.
We present an alternative picture that allows SN~1998bw to have
an explosion energy and ejecta mass consistent with 
core collapse supernovae although at the bright end of the typical range.
  We specifically
propose that all SN~Ic are significantly asymmetric and that 
SN~1998bw is a SN~Ic that is distinguished principally by being 
viewed close to the symmetry axis.

We investigate the hypothesis that SN~Ic and SN~1998bw are
the result of an aspherical explosion along the rotational axis of a 
basically spherical, non-degenerate C/O core of a massive star.
Light curves for aspherical explosions are computed  assuming 
an ellipsoidal geometry for the ejecta. Guided by the polarization 
observations of ``normal" SN~Ic and related events, we assume an axis ratio
of 2 near maximum light. The evolution of the iso-density
contours with time is discussed.
We show that the light curve of SN~1998bw may be understood 
with an explosion energy of $2\times10^{51}$ ergs, a total ejecta
mass of 2 $M_\odot$, and a mass of $^{56}Ni$    of 0.2\m\  
if it is observed at a large angle ($\geq 60^o$)
with respect to the plane of symmetry. In this picture, the high expansion 
velocities are a direct consequence of an aspherical 
explosion mechanism which, in turn, produces oblate iso-density contours. 
Prolate iso-density contours are ruled out.
This interpretation suggests that the fundamental core-collapse 
explosion process itself is strongly asymmetric.

\end{abstract} 
 

\section{Introduction}
 
In the year and a half of great excitement following the discovery
of the first optical counterparts of \grbs, one of the most 
interesting developments was the detection of the \grb\
GRB~980425 by BeppoSaX (Boella et al. 1997) and BATSE (Fishman et al. 1993).
Due to its correlation in time and location, this \grb 
 has a 
high probability of being associated with SN~1998bw (Galama et al 1998).
This connection is supported by the association of
a relativistically expanding radio source with
SN~1998bw (Kulkarni et al. 1998). 
From optical spectra, SN~1998bw was classified as a 
SN~Ic by Patat and Piemonte (1998).   SNe~Ic 
are commonly attributed to the explosion of the non-degenerate
C/O cores of massive stars that have lost their hydrogen-rich and 
helium-rich layers by the time of the explosion 
(see e.g. Iwamoto et al. 1994; Clochiatti \& Wheeler 1997).
 
Consistent with typical SNe~Ic, the light curve of SN~1998bw 
showed a fast early rise and reached
a peak brightness of $13.6^m$ in V $\approx$ 17 days 
after the explosion.  After maximum, SN~1998bw showed an 
exponential decline typical for a supernova light curve, 
but unlike the afterglows that have been observed
in other \grbs.
Although the data base for SNe~Ic is rather sparse, 
SN~1998bw appears to be unusual.
What sets SN~1998bw apart from most of the other observed SNe~Ic 
is the large intrinsic brightness and higher expansion 
velocities as indicated by the Si II and Ca H and K lines
(30 to 50 percent higher at maximum light than  SN1994I and SN1983V,
Clochiatti \& Wheeler 1997).
For a detailed presentation of the observational data see 
Galama et al. (1998). In addition, linear polarization
has been reported by Kay et al. (1998) at the level of 0.5 \% on 21 June,
about two months after the explosion. 

The luminosity of SN~1998bw can be inferred from the redshift 
of the host galaxy and its reddening.  The redshift 
of the host galaxy ESO 184-G82 is very well known,
$z=0.0085 \pm 0.0002 $ (Tinney et al. 1998). The host galaxy 
is a member of the group DN 193-529 (Duus \& Newell 1977).
Galama et al. (1998) inferred the Galactic foreground extinction 
to be $A_V = 0.2^m$ from a combination of Cobe/DIRBE  
and IRAS maps (Schlegel et al. 1998). Assuming a Hubble constant
of 67 km s$^{-1}$ Mpc$^{-1}$, the distance can be derived to be 36 Mpc. 
The main sources of uncertainty in the
luminosity determinations are the following: a) The redshift 
corresponds to a cosmological expansion $v_z$ of 2550 \kms.  This
means the group is not yet fully in the Hubble flow, and that 
peculiar velocities of the group and of the galaxies within
the group may be significant, of the order of 10 to 20 \%
of the deduced redshift velocity. 
b) There is an uncertainty in $H_0$ of about 10 \%.
c) Fluctuations in the foreground extinction may be of the 
order of $0.1^m$ and there may be some extinction is the host galaxy.
From a) and b), the distance of SN~1998bw with appropriate
uncertainty can be estimated to be $36 \pm 7$Mpc. 
With the addition of the uncertainty in
the reddening, this corresponds to an uncertainty of 
$\pm 0.5 $ magnitudes in the intrinsic brightness.
 
Even within this uncertainty, SN~1998bw appeared to be a 
very luminous event compared to ``normal" SN~Ic,
favoring the notion that SN~1998bw was a ``hypernova" event 
(Paczy\'nski 1997) with ejected masses and explosion energies
at least a factor of 10 larger than other SN~Ic. 
Based on their light curve calculations,
Iwamoto et al. (1998) found good agreement (better than $0.3^m$ over 40 days)
with the observed bolometric light curve for models based on the explosion
of C/O cores with ejecta masses between 12-15 $M_\odot$, 
final kinetic energies of $E_{kin}\approx $ 20 to 50 foe 
(1 foe = $10^{51}$ erg),  
and the mass of $^{56}$Ni between 0.6 and 0.8 $M_\odot$. As 
noted by the authors, one problem with their calculation
may be that the bolometric correction was assumed to be constant with time.
Another problem is  that the line width of the synthetic 
spectra are too narrow by a factor of 2 to 3 compared to the observations.
This indicates that in the models the absorption is formed over too 
narrow a region in velocity space. This implies that over the phase
considered the photosphere in the homologously expanding envelope extends
over too narrow a radial region.  One interpretation of the narrow 
theoretical absorption features is that the envelope masses are
too large.  Woosley, Eastman \& Schmidt (1998) found best 
agreement with a ``hypernova" model that ejects a C/O envelope of 
about $ 6 M_\odot $ with $E_{kin}= 22~foe$, and $M_{Ni}$ of 0.5 $M_\odot$. 
For this model, the bolometric and monochromatic light curves 
differ from the observations by 0.5 to 1 magnitude over the course  
of 20 days and all the computed color indices (B-V, V-R, V-I)
are too red by about the same amount.
Despite the uncertainties and disagreements in detail, the common 
denominator of the calculations by Iwamoto et al. and Woosley et al.
are  explosion energies that are larger by 
a factor of more than 20 and $Ni$ masses that are 
larger by a factor of 5 to 10 compared to typical core collapse
supernovae. This clearly points  towards  the need for a 
``hypernova" scenario.
 
Guided by the deduced properties of more traditional core 
collapse supernovae and SNe~Ic in particular, 
we seek to demonstrate that an alternative explanation 
is possible for which SN~1998bw falls in the range of ``normal" SNe~Ic
parameters.
 
Although neutron stars are observed throughout the Galaxy
and there is general agreement that SNe~II and SNe~Ib/c  
are the result of a core collapse, the nature of the 
explosion mechanism remains unsolved. Both spectral analyses and 
light curve calculations support the picture that
SNe~II, SNe~Ib and SNe~Ic may form a sequence involving core 
collapse within a massive H-rich envelope,
within a core denuded of most of its hydrogen but which retains
substantial helium, or within a C/O core which has lost most of its 
helium, respectively.
The analysis of spectra and light curves gives essentially no
insight into the geometry of the expanding envelope. 
It has been demonstrated that, on the level of current 
spectral fitting, spectra produced in aspherical models can be 
mimicked by spherical calculations (H\"oflich, Wheeler, Hines \& Trammel 1995).
Polarization, however, provides a unique tool to explore asymmetries.
As with so many issues, SN~1987A represented a breakthrough in this area
by providing the first detailed record of the spectropolarimetric 
evolution that led to the conclusion that the envelope of SN~1987A 
was aspherical by about 10 \% (M\'endez et al., 1988, 
H\"oflich 1991, Jeffery 1991). In the following years,
similar amplitudes of polarization have been observed in a 
program at McDonald Observatory. These observations continue to
confirm the early qualitative conclusion that all SNe~II 
are polarized at about this level (Wang et al. 1996, Wang, Wheeler \& H\"oflich
 1998).  
There can be a number of reasons why light from a supernova
is polarized (Wang et al. 1996, 1998)
with an intrinsically strongly aspherical explosion
in the core region only one possibility.
There is a trend, however, for the observed polarization 
to increase in core-collapse supernovae with decreasing envelope mass,
e.g. from SN~II to SN~Ic (Wang et al. 1998).  
For SN~1993J, in which only a small
mass hydrogen envelope remains, the observed linear polarization 
was as high as $\approx 1.0$ to $1.5 \% $ (
Trammell, Hines \&  Wheeler 1993; Tran et al. 1997). 
For the SN~Ic 1997X the polarization was even higher 
(Wang \& Wheeler 1998). 
This trend, while tentative, clearly points toward the 
interpretation that the explosion itself is strongly asymmetric
and that this fundamental asymmetry is revealed more
clearly as more envelope matter is removed. 
 
 \begin{figure}[t]
  \psfig{figure=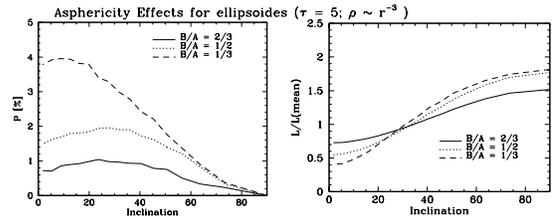,width=8.0cm,rwidth=7.0cm,clip=,angle=270}
 \caption
 { Polarization (left) and luminosity (right) as a function
 of the angle of the line of sight with respect to the equator are shown for oblate ellipsoids with
 various axis ratios. The energy source is assumed to be at an optical depth
 of 3 in Thomson scattering.}
 \end{figure}
 
From theoretical calculations for scattering dominated atmospheres, 
this size of polarization, $\gta 1 $ \%,
requires axis ratios of the order of 2 to 3, 
making these objects highly aspherical. The 
consequences for the dependence of the apparent 
luminosity $L(\Theta )$ on the angle $\Theta$  of the line of sight has been 
pointed out and studied in detail (H\"oflich 1991, H\"oflich et al. 1995a).  
In Fig. 1, we show the dependence of the luminosity and 
polarization for oblate ellipsoids with axis ratios between 1.5 to 3
for  power law density profiles $\rho \propto r^{-3}$. 
In this example, the energy source is assumed to be located 
at an optical depth of 5. The angular dependence of 
$L(\Theta)$ changes the apparent luminosity by almost a factor of 4. 
For a systematic study of the influence of the optical depth, 
the geometries, the density structure, and the geometry of the envelope, 
 see H\"oflich (1991, 1995a).
 
Given the ubiquitous presence of polarization in core collapse
supernovae and SN~Ic,  inclusion of asphericity effects in SN~Ib/c may 
prove to be critical to their understanding
(Wang et al. 1998). The same may be true for SN1998bw.
 
In this work, we present a first approach to the problem of asymmetric 
light curves for SN~Ic and SN~1998bw.
In the next section, the basic concept and methods are outlined. 
General results for various geometries are discussed in 
\S 3. In \S 4, our results are compared 
to the observations.
A final discussion, conclusions are presented in \S5.
 
\newpage

\section{Description of the Concept and  Numerical Methods}
 
\subsection{General Concept}

For the initial setup, we use the chemical and density 
structures of spherical exploding C/O cores of Nomoto and Hashimoto (1988). 
These structures are scaled homologously to give models of
different ejecta mass. This is an approximation, but the details 
of the chemical profiles are not expected to effect the light 
curves. 
The explosion energy is deposited in the form of thermal energy 
at the inner edge of this spherical model. 
The mass cut of the core 
is adjusted to provide the ejection of the desired $^{56}Ni$ mass.
A problem may be the final  distribution of $^{56}Ni$ which
is fundamentally uncertain in the absence of an understanding
of the core collapse and fall-back processes. Possible effects of
altering the structure of the $^{56}Ni$  regions is discussed in \S 4.
A more important approximation is that we assume that
iso-density contours are ellipsoidal.   The actual structures
resulting from strongly aspherical explosions may be significantly
different (see \S 4, 5).

The hydrodynamical explosion and the spherical light curves are 
first calculated (\S 2.2).  Aspherical envelopes are constructed by assuming 
directionally-dependent expansion ratios. An appropriate 
redistribution of the energy is calculated to construct the aspherical 
light curve (\S 2.3).
 
\subsection{The Explosion and Light Curve for Spherical Geometry}
 
Spherical explosion models are calculated using a one-dimensional
radiation-hydro code (H\"oflich, M\"uller \& Khokhlov 1993,  
H\"oflich \& Khokhlov 1996) that solves the hydrodynamical 
equations explicitly by the piecewise parabolic method 
(Colella and Woodward 1984).
The models use 234 depth points. The
explosion energy is deposited as thermal energy over the inner 10 
depth points.

The spherical code also simultaneously solves for the radiation transport.
The radiation transfer portion of the code
consists of (i) an LTE radiation transfer scheme 
based on the time-dependent moment equations which are
solved implicitly, (ii) a detailed equation of state with an elaborate
treatment of the ionization balance and the ionization energies, (iii)
time-dependent expansion opacities  which take into account the
composition structure of the explosion model, and (iv) a Monte Carlo
$\gamma$-ray deposition scheme which takes into account all relevant
$\gamma$-ray transitions and interaction processes 
(H\"oflich, Khokhlov \& M\"uller 1992).

The energy and radiation momentum equations with variable 
Eddington factors are solved in the comoving frame (Mihalas, 1978).
The energy equation is solved with appropriately frequency-weighted
mean opacities (H\"oflich et al. 1993).
The Rosseland, energy and Planck mean opacities are obtained
from the calculated monochromatic expansion opacities (see below),
At each time step, we use $T(r)$ to determine the
Eddington factors by solving the frequency-dependent 
radiation transport equation in the comoving frame in about 
100 frequency bands and integrate to obtain the 
frequency-averaged Eddington factors.  We  use the
latter to iterate the solution with the frequency-integrated
energy and flux equations (H\"oflich, Wheeler \& Thielemann 1998).

In rapidly expanding atmospheres, the opacity, $\kappa_{\nu}$, can be
significantly enhanced due to line blocking effects as different atomic
transitions are Doppler-shifted into the frequency of a propagating photon.
Because in supernovae the Doppler shift due to the velocity
field is often much larger than the intrinsic line width, the influence of
lines on the opacity can be treated quite accurately in the Sobolev
approximation (Sobolev 1957, Castor 1974, H\"oflich 1995b).
The corresponding absorption probability is calculated, 
including bound-bound, bound-free and free-free opacities
based on  data for the atomic line transitions from 
Kurucz (1995)  and  Cunto and Mendoza (1992). 
Typically, frequency averaged opacities
are of the order of 0.05 g cm$^{-2}$. 

Both Thomson scattering with cross section, 
$\sigma _T$, and line scattering are taken into account.  
For line scattering, an estimate of the fraction of incident photons
that are thermalized or otherwise redistributed to other frequencies
is crucial.  An equivalent two-level formulation for the source
function is adopted with
parameters determined by comparison with detailed NLTE-calculations 
that properly take into account the photon redistribution and 
thermalization processes (H\"oflich 1995b).
 
To calculate the monochromatic light curves, we use the $T(r)$ 
with the time dependence of the structure given by the frequency-integrated 
solution of the momentum equations.  This evolving temperature
structure is used to solve  the frequency-dependent transfer equations 
in LTE every  0.5 days. This co-moving solution is used to
compute $L_\nu$ in the observer's frame. 
The broad band light curves are determine by convolution 
of $L_\nu$ with the filter functions.
Here, we use a few hundred frequency bands.
 
\subsection{Light Curves for Aspherical Geometry}
 
 Aspherical density structures are constructed from the 
spherical density distribution.  The simple models presented
here are based on the assumption that there is a strongly
aspherical explosion in the approximately spherical core.
We have not done the actual hydrodynamics to produce such
an asymmetric configuration, a task we postpone to future
work.  Here we impose the asymmetry after the ejecta has
reached the homologous expansion phase.  We generate an
asymmetric configuration by preserving the mass fraction
per steradian from the spherical model, but imposing a different
law of homologous expansion as a function of the angle $\Theta$
from the equatorial plane.  By neglecting the actual hydrodynamics,
we assume there is no redistribution of material in the
$\Theta$ direction. 
Note that, for typical density structures, a higher energy deposition 
along the polar axis results 
in oblate density structures with prolate Lagrangian surfaces (see below).
Such an energy pattern may be produced if jet-like structures 
are formed during the central core collapse as suggested by 
Wang \& Wheeler (1998). In contrast, a prolate density structure
would be produced if more energy is released in the equatorial region
than in the polar direction.
Aspherical iso-density structures could also be produced if the 
original structure were aspherical i.e. if mass is distributed 
non-uniformly in $\Theta $ prior to the explosion 
compared to a spherical density configuration, by, for instance, rotation.   
We regard this possibility as less likely to produce large 
asymmetries since transverse pressure gradients during the explosion 
will tend to make originally aspherical mass distributions 
more spherical as the expansion proceeds ( Chevalier \& Soker 1989).

The initial density structure is spherical, i.e.
$$\rho(R, \Theta ) = \rho (R), \eqno{(1)}$$
\noindent
with R the initial distance of a mass element from the center.
Homologeous expansion of the spherical density distribution is assumed
with a scaling that depends on angle, i.e., 
$${v(R,\Theta) \over R} = C(\Theta),\eqno{(2)}$$
\noindent
so that
$$r(\Theta ) =  C(\Theta ) ~R ~t,\eqno{(3)}$$
\noindent
where t is the time since the explosion
and $ r(\Theta)$ is the distance of the 
mass element after time t. For the spherical configuration,  $C=v_{sphere}/R$.
 
Because little is known about the general geometry of the envelopes,
we assume ellipsoidal iso-density contours with an axis ratio 
E= B/A of the photosphere at a reference time
 where A is the distance of the photosphere  in the x-y (symmetry) plane and 
B is the distance in the (axial) z-direction.  
This  contour is given by
$$ r(\Theta) = r(\Theta = 0)~{\sqrt {(cos^2(\Theta ) + 
    E^2 ~ sin ^2(\Theta ))}}.\eqno{(4)} $$
The homology scaling constant, $C(\Theta)$, is then 
chosen to produce the desired axis ratio starting from the spherical configuration.
This means that $C(\Theta)$ has the distribution 
$$C(\Theta )= C(\Theta = 0) *\Biggl({\rho (\tilde R(\Theta)) 
   \over  \rho (\tilde R(\Theta = 0))}\Biggr)^{1 \over 3 }. \eqno{(5)}$$
 Here $\tilde R(\Theta)$ denotes the distance from the center of the original spherical model 
at the beginning of homologous expansion of the mass element that is mapped to angle $\Theta $
at the time the iso-density contour is defined. We take this time to be 20 days after the explosion,
 near maximum light.
 In our actual calculations, we use Eq. 5 and do an implicit iteration to 
determine the functions $\tilde R(\Theta)$ and $C(\Theta)$.
 We note that the solution of Eq. 5 becomes independent from the time when the homologous expansion is
establish as long as $R << r(20d)$.
 For power law density profiles ($\rho \propto r^{-n}$) with $n \neq 3$, 
the distribution of $C(\Theta)$ with  angle would be 
$$C(\Theta )=C(\Theta = 0)\Biggl(\sqrt{cos^2(\Theta)+E^2 ~sin^2(\Theta)} \Biggr)^{-n \over 3-n}, \eqno{(6)}$$
\noindent
and that of $\tilde R(\Theta ) $ would be
$$\tilde R(\Theta )=\tilde R(\Theta = 0)\Biggl(\sqrt{cos^2(\Theta)+E^2 ~sin^2(\Theta)} \Biggr)^{3 \over 3-n}. \eqno{(7)}$$
 In general, $C(\Theta)$ depends on the underlying density structure as
the density gradient (with the equivalent, local slope $\tilde n$) changes between $\tilde R(0^o)$ and 
$\tilde R(90^o)$. In all models discussed here,  Eq. 6 is very close to the exact
solution as $\tilde n$ varies little over the photosphere.  For example, 
 the photosphere  is formed around mass fraction 0.45 in our reference model
where $\tilde n$ is nearly constant (Fig. 2).
 
The total explosion energy is normalized to that of the 
corresponding spherical model, i.e.
$$ \int_{V} \rho (R) \times v_{sphere}^2(R) dV = \int_{V} \rho(R) \times v^2(R,\Theta) dV \eqno{(8)}$$ 
where V is the volume containing the ejecta.
 This determines $C(\Theta = 0)$.
For mean density slopes $<  3$, oblate structures are produced
if more energy is released in the polar direction. 
The energy release required to produce an aspherical configuration 
formally diverges for $n \rightarrow 3 $.
 
 Because of changes in the density structure, iso-density contours within and beyond the
photosphere at day 20 will be nearly, but not exactly ellipsoidal (see below and Figs. 2, 4 \& 5). 
 As the density gradient (or $\tilde n$) changes, the axis ratio will change with time 
and mass coordinate. For the cases discussed 
in this work, the variations in geometry remain small except in
the outer layers in which the photosphere
is formed well before maximum light (see below).

We use a Monte Carlo code to compute the light curve of
the asymmetric configuration.
 This code is capable of handling arbitrary three-dimensional geometries, 
both for the density and the distribution of the sources. 
 The background is assumed to be stationary.
In the calculations, the angular space of the emitted photons is 
discretized by 60 zones in the $\Theta $ direction.               
The extinction $k$, opacity $\kappa $ and the emissivity of the photons $\eta$ are assumed to be   
constant on iso-density contours at each time t. The values of $\kappa $,$k$, and $\eta $ are taken to  
be the values of an equivalent mass faction at radius R in the spherical model. The equivalent radius R
is defined such that $R=C <v>$ where C is the homology constant in the spherical model and
$<v>$ is the surface-area weighted average of the velocity over the iso-density contour. The values
of $\kappa $, $k$, and $\eta $ are thus given by  
$$ \kappa (\bar R(r,\Theta,t),t) = \kappa_{sphere} (R,t),\eqno (9) $$ 
$$ k(\bar R(r,\Theta,t),t) = k_{sphere} (R,t) \eqno (10) $$ 
and
$$ \eta(\bar R(r,\Theta,t),t) = \eta_{sphere}(R,t) \eqno (11) $$ 
where $\bar R(r,\Theta,t)$  is the general mapping function of mass elements from the spherical model
 to respective iso-density contours, e.g. $\tilde R (\Theta) = \bar R (r_{photosphere}, \Theta, 20 days)$.
 Here, we  take the local quantities from  a representative mass element of the spherical model.
 Note that the mass elements on the iso-density contours arise from different
depths in the original spherical model with varying $k$, $\kappa $  and $\eta$. This forbids a direct
mapping of the properties of the  lagrangian mass elements because it would strongly violate 
the energy balance between absorption and emission. 
 Photons are generated in the Monte Carlo code in proportion to $\eta$.
The result of the Monte Carlo calculation is the relative distribution of the
luminosity as a function of angle and time, $l (\Theta, t)$.
 The influence of the distribution functions of the photon sources,  $\eta$, for
various density structures has been studied in detail (H\"oflich 1991). From these studies,
errors in $l (\Theta, t)$ can be expected to be $\approx $ 10 \%. For more details, see H\"oflich (1991, 1995a)
and H\"oflich et al. (1995).

The bolometric and broad band light curves are constructed 
by multiplication of the spherical light curves at each time step by 
 the redistribution factors $l(\Theta, t)$
calculated  by the Monte Carlo code, i.e. 
 $$L_{bol,BVRI}(\Theta,t) = l(\Theta, t) \times L_{bol,BVRI}(mean), \eqno{(12)}$$
where $L_{bol,BVRI}(mean)$ stands for the bolometric and BVRI light curves for
spherical geometry. 
 
 In Fig. 5, we show the change of the axis ratio of the iso-density contours at the photosphere 
with time. The  variations are rather slow. Therefore,
to  first order, variations of the geometry of the envelope can be neglected
over  the diffusion time scale of a photon.
Typical conditions at the photosphere and therefore the colors 
are expected to be similar in both the spherical and aspherical models.
For an oblate ellipsoid with an axis ratio of 2, 
the area seen pole-on is about twice as large as the area 
of the corresponding spherical structure and the value of $L/L(mean)$ 
observed along the symmetry axis is also about 2 (see Fig. 1)
so the flux is about the same. Quantitatively, 
the energy flux $F(\Theta)$ is found to be similar 
in the spherical and aspherical configurations to 
within $\approx 40 \% $. 
To first order (Wien's limit), a change of $F(\Theta)$  by 40 \% 
corresponds to a change in color indices  by about 0.1$^m$.
 
Eq. 12 contains the implicit assumption that the
 mean diffusion times are the same for spherical and aspherical envelopes.
In reality, the diffusion time scale will be shorter for aspherical 
configurations because, at a given mass element,
the diffusion time varies as $r^{-4}$ because of its quadratic dependence on 
the optical depth, whereas the density goes as $r^{-3}$.
The change in the diffusion time scale mostly effects the very early phases of the light curve 
when it is much longer  than the expansion time scales. During the initial rise, 
the luminosity will be underestimated (see \S4).
This represents another approximation that is consistent with these trial
calculations, but inadequate for a full understanding of
the radiative transfer in asymmetric configurations.  
We intend to do full aspherical radiation-hydro calculations
that will eliminate this mapping process.
 
All models in the next section are constructed in such a way that, at day 20,
the axis ratio at the photosphere is 0.5 and 2.0 for oblate and 
prolate ellipsoids, respectively.
The photosphere is defined at the density at which 
the optical depth in Thomson-scattering equals 2/3 in the spherical model. 
In comparison to the spherical model, the expansion parameters are 
a factor of $\approx 2.2 $ larger along the pole for oblate ellipsoids 
and a factor of $\approx 1.5$ larger for prolate ellipsoids 
along the equator.

 \begin{figure}[t]
  \psfig{figure=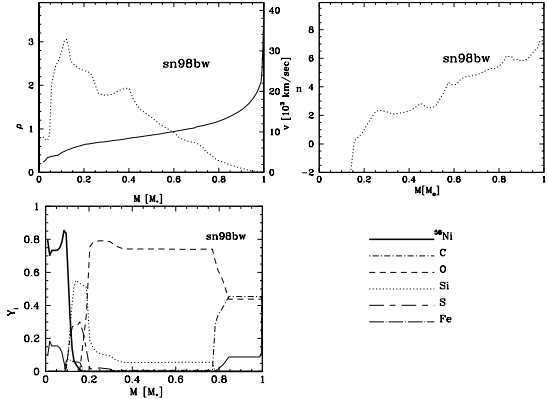,width=8.0cm,rwidth=7.0cm,clip=,angle=270}
 \caption
 { Density and velocity structure after the explosion (upper left) 
 and chemical structure (lower left) as a function of envelope mass.
 In the upper right plot, the power law
 index of the density profile as a function of mass is given.
  The total mass of the ejecta of our reference model
 is 2 $M_\odot$ and the kinetic energy is $2\times10^{51} erg$. }
 \end{figure}
 \begin{figure}[t]
  \psfig{figure=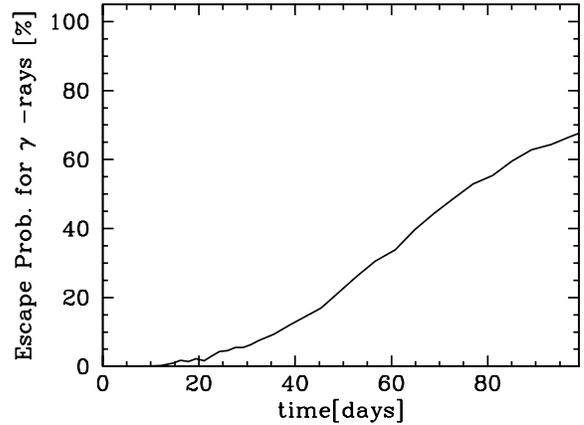,width=8.0cm,rwidth=6.0cm,clip=,angle=270}
 \caption
 {Escape probability for $\gamma $-rays as a function of time for the model of Fig. 2.}
 \end{figure}
 \begin{figure}[t]
  \psfig{figure=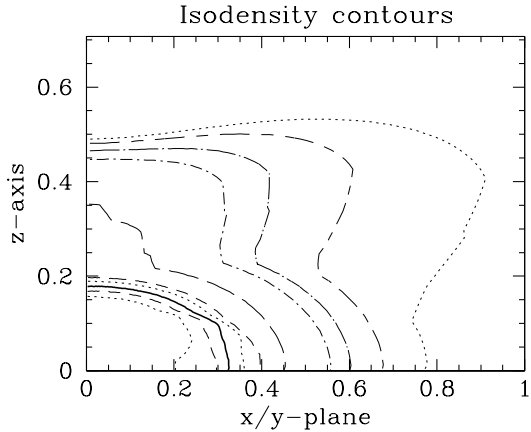,width=8.0cm,rwidth=7.0cm,clip=,angle=270}
 \caption
 { Iso-density contours for the reference model with oblate geometry. The contours are defined 
 by the location in the equatorial plane of mass elements that 
 corresponding to mass fractions in the spherical model of
  0.1, 0.3, 0.5 (thick line), 0.6, 0.7, 0.8, 0.9, 0.93, 0.96, 0.98
 The iso-velocity contours are
 spheres and Lagrangian mass elements are prolate as given by Eq.5 and 6.
 }  
 \end{figure}
 \begin{figure}[t]
  \psfig{figure=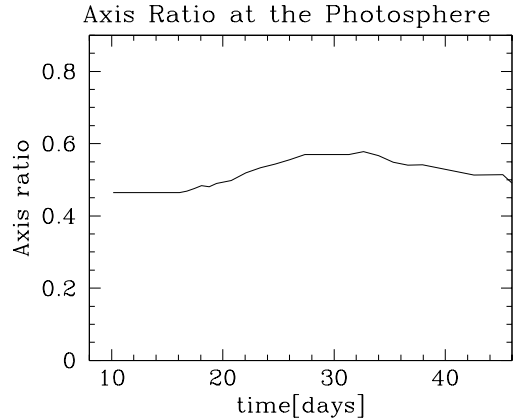,width=8.0cm,rwidth=7.0cm,clip=,angle=270}
 \caption
 {The
  axis ratio of the iso-density contour at the photosphere ($\tau_{scattering,\Theta=0^o}$ = 0.667) for the
 oblate ellipsoid is given as a function of time for the reference model  with  oblate geometry.
  The axis ratio varies with time because the power law
  index n varies with depth.}
 \end{figure}

\section{Results}
 
We first calculated aspherical light curves based 
on C/O cores with initial total masses of 1.8 and  2.1 $M_\odot$ 
(models CO18 and CO21) 
that have ejecta masses of $0.5$ to $0.8 M_\odot $ (CO18 and CO21, 
respectively), explosion energies of $E_{kin} = 10^{51} $erg , 
and that eject $\approx 0.07 -0.08~ M_\odot $ of $^{56}Ni$.
In previous work, model CO21 was found to give a good representation of the 
BVRI light curves of the SN~Ic 1994I (Iwamoto et al. 1996). 
This first set of models failed to represent SN~1998bw in several respects.
The luminosity is increased by about a factor of 2 in the polar
direction for the oblate model and about a factor of 1.5 in
the equatorial plane for the prolate model.
Despite this boost in the luminosity, 
these models are too dim even within the uncertainties discussed
in the introduction. The maximum was also too early by about 5 days. 
Finally, these models are very blue at maximum light. 
An increase of the asphericity does not solve these problem (see Fig. 1).

To boost the total luminosity to the level of observation 
we increased the amount of ejected  $^{56}Ni$ to 0.2 $M_\odot$. 
This quantity of nickel is still below the estimate 
of 0.3 $M_\odot$ in SN1992 which is the most luminous 
SN~II event observed so far (Schmidt et al. 1997) and substantially less than 
in the ``hypernova" models of Iwamoto et al. (1998) and Woosley et al (1998).
The time of maximum light is 
rather insensitive to asphericity effects (Figs. 7 \& 8).
In principle, the rise time could be increased by an increase in mass or,
alternatively, by a decrease in the expansion rate.
The latter option can be ruled out because of the need
for large expansion velocities. The need to delay the time to maximum
therefore suggested the need to increase the ejecta mass with
an appropriate increase in the kinetic energy to provide the
observed expansion.
 \begin{figure}[t]
  \psfig
  {figure=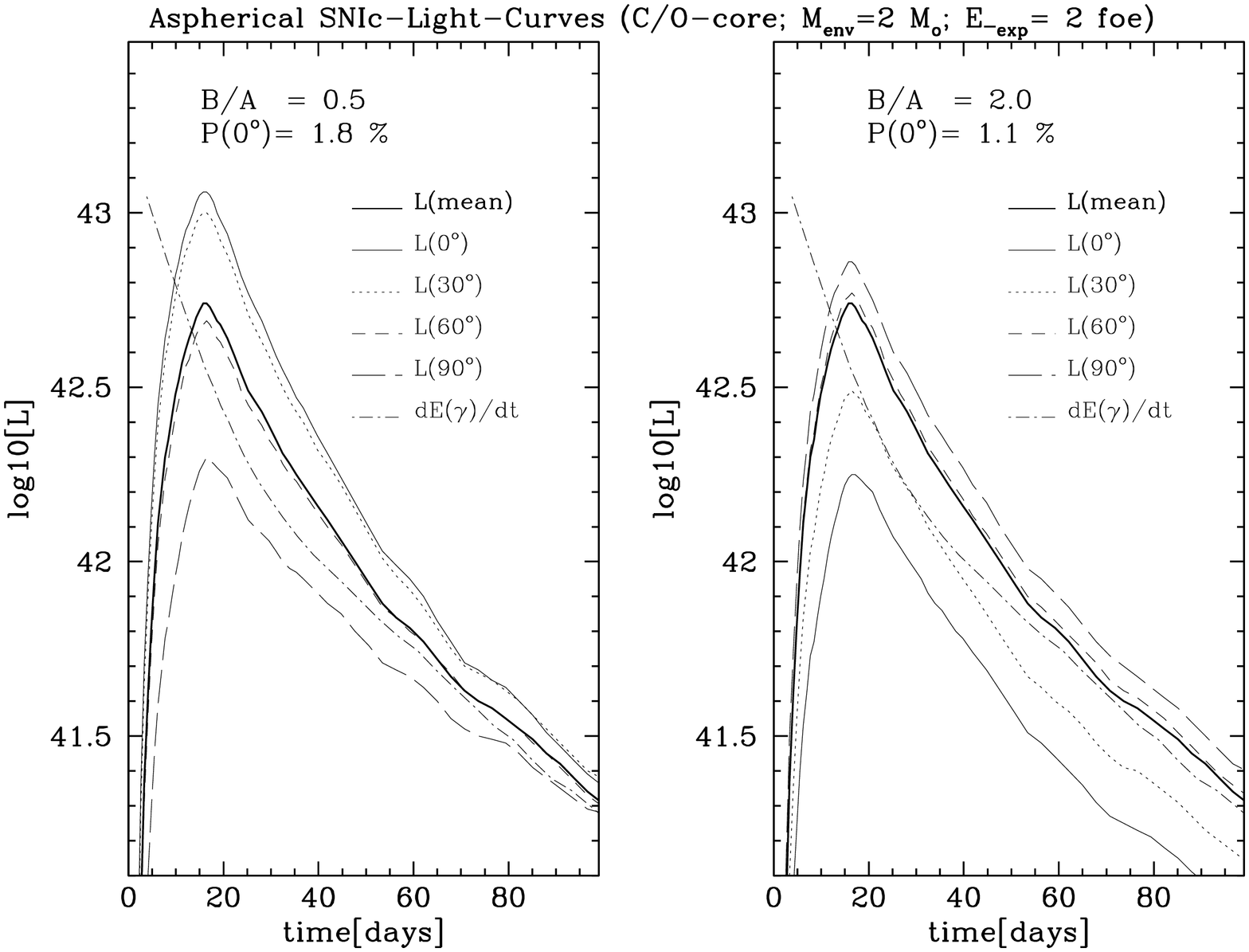,width=8.0cm,rwidth=7.0cm,clip=,angle=270}
 \caption
 { Directional dependence of the bolometric light curve for oblate (left) and prolate
 (right) ellipsoids for the reference model.
  The luminosity of the corresponding spherical model is shown as $L(mean)$.
 In addition, the instantaneous \gr\ deposition is shown. $P(0^o)$ is the polarization at 
 maximum light ($ P(\Theta) \approx P(0^o) \times cos^2\Theta$).
 }
 \end{figure}
 
 \begin{figure} [t]
  \psfig{figure=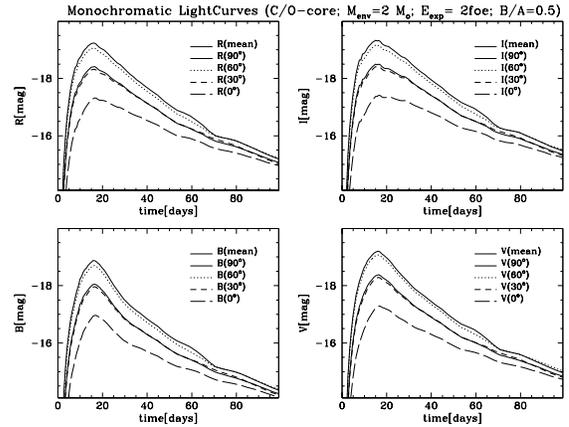,width=8.0cm,rwidth=7.0cm,clip=,angle=270}
 \caption
 { Same as in Fig. 6  but for the broad-band light curves
  and the oblate geometry only.}
 \end{figure}

We thus computed a series of models with $M_{ej}=2 M_\odot$,
$E_{kin}=2\times10^{51}$ erg and $M_{Ni} = 0.2$\m. 
The chemical structure and the final density structure
during the phase of homologeous expansion are given in Fig. 2. 
The escape probability for $\gamma $-rays from the radioactive decay 
of $^{56}Ni$ and $^{56}Co$ increases rapidly 
after maximum light (Fig. 3), with consequences as discussed below. 
 
For illustration, the iso-density contours  are given for the oblate structure in Fig. 4.
They are close to elliptical except for the very outer layers. 
 Note that shape of the  photosphere and the corresponding change 
in the polarization early on provides a sensitive tool to study details of the 
asphericity in the distribution of the explosion energy.
 
The density slope at the photosphere changes with time. 
This causes the axis ratio at the photosphere to vary.
As an example, we give the time evolution of the oblate ellipsoid 
axis ratio in Fig. 5.  The envelope becomes transparent at about day 70.
 
In Fig. 6, $L(\Theta)$ is given for oblate and prolate ellipsoids. 
Asphericity of the amplitude we have assumed here
can change the luminosity over a range of roughly 2 magnitudes,
depending on the line of sight.
For oblate ellipsoids, the luminosity is
enhanced pole-on whereas for prolate structures 
the enhancement occurs in the equatorial direction. 
Combined with the polarization properties, 
this provides a clear separation between oblate and prolate geometries. 
The linear polarization $P$ always goes to
zero if the structure is seen pole-on and  $P$  
increases for lower latitudes (H\"oflich 1991). 
Statistically, for oblate density structures we thus expect 
that high apparent luminosity will be associated with low 
polarization and low luminosity will be associated with
high polarization. The opposite trend is expected for prolate geometries. 
The maximum amplification of L is larger in oblate structures 
compared to prolate configurations because of the differences
in the projected area for a given axis ratio.

 Observations of the polarization of SN~1998bw 23 days after
the explosion show little polarization ($\leq  1 \% $, Patat et al. 1998). 
By day 58,
the intrinsic polarization was reported to be 0.5 percent (Kay et al. 1998).  
Polarization data on SN~Ic are rare, but this value is less than observed in 
SN~1993J and in SN~ Ic 1997X (Wang et al. 1998, Wang \& Wheeler 1998). 
SN~1998bw was also rather bright.  The polarization
is time dependent, and we do not necessarily know the maximum
value in SN~1998bw.  Nevertheless,  the evidence suggests that, 
if asymmetry is involved, oblate
geometries are to be favored over prolate geometries.
We therefore concentrate on oblate structures in the following.

The corresponding broad-band light curves for the  oblate geometry 
are  shown in Fig. 7.  The absolute magnitude of 
our model with $M_{ej}=2 M_\odot$ $E_{kin}=2\times10^{51}$ erg 
and $M_{Ni} = 0.2$\m\ can be as high as $-19.0, -19.3, -19.4 $ and
$-19.5^m$ in B,V,R and I, respectively, when viewed pole-on. 
The decline rate tends to decrease from the IR to B.
 
\section{Theory vs. Observations}
 
For the comparison between the observed and theoretical light curves, 
we assumed an interstellar extinction 
$A_V=0.2$ and a distance of 36 Mpc. 
We have used the relative calibration of Woosley et al. (1998)
for the ``bolometric light curve" that was obtained by integrating over 
the UBVRI photometry.  The broad-band data was obtained from 
Galama et al. (1998).  In the following, it should be remembered 
that the absolute calibration of the light curve is uncertain
by a factor of $\pm$ 40 to 50 percent, as shown in the introduction. 
 
The comparison of the model and observed bolometric light curve 
in Fig. 8 shows that for our model, SN~1998bw 
must be obsreved from an angle of $\ge 60^o$ from the equator. 
The same conclusion can be tentatively drawn from the small size of the
linear  polarization observed.
Within the framework of our approach, any prolate density structure 
can be ruled out.  The required boost in the luminosity by
a prolate geometry would require even larger asphericities, 
but the required asphericity would imply linear polarization
of several percent (H\"oflich 1991).
 \begin{figure}[t]
  \psfig{figure=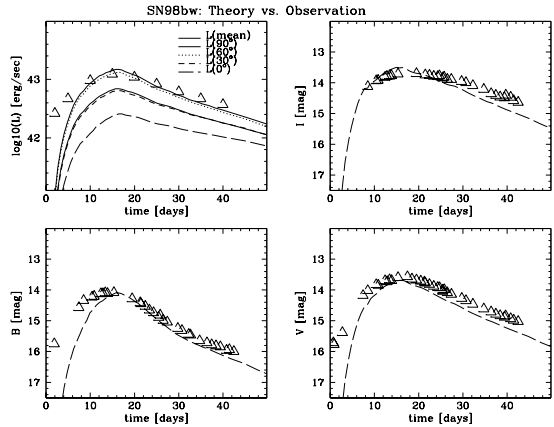,width=8.0cm,rwidth=7.0cm,clip=,angle=270}
 \caption
 {Comparison of bolometric and broad-band  light curves
  as observed for SN~1998bw with those of an oblate ellipsoid seen at high angle with respect to the equator.}
 \end{figure}
 
 Overall, the broad-band light curves agree with the data within the 
uncertainties expected from our approximations. 
In particular, the time of maximum agrees within 2 days except in 
the I band.
The intrinsic color excess B-V matches the observations within $0.1^m$ 
and after the initial rise of $\approx 7$  days, the agreement in the colors 
is better than $0.3^m$. 
The main discrepancy with the observations occurs during 
the first week of the rise when our models are too dim.
During this phase, the diffusion time scales are
much longer than the expansion time.  Under these conditions, 
our approximation of redistribution of the energy
of a spherical model to the ellipsoidal configuration 
breaks down. The diffusion time scale 
goes with the square of the optical depth so the diffusion
time on the rise, and hence the luminosity are probably
especially uncertain. 

The decline after maximum is slightly too steep in the models
both in the bolometric and broad-band light curves. 
This is likely to be related to the energy generation in the 
envelope by \gr ~deposition or to the change in the escape 
probability of low energy photons.  The 
decline rate after maximum light might be affected by 
several factors.
The instantaneous radioactive decay input becomes a more 
dominant contribution at later times as previously stored thermal
energy is dissipated. The decline rate immediately after peak 
will thus be reduced by increasing the amount of radioactive $^{56}Ni$. 
By day 40 to 50, the diffusion time scales are only a few days.
The increase of $^{56}Ni$ needed to fill the gap between the bolometric 
(and broad-band) light curves is estimated to be $\approx 40 \%$. 
Another, related, possibility is suggested by Fig. 3. 
The escape probability for $\gamma $-rays increases rapidly between
day 20 and 80. Variations in model parameters may shift 
the time of this rapid change to earlier or later epochs.
The first option seems to be unlikely as it implies that the total luminosity
at later times should mainly be provided by 
the energy deposition due to positrons. 
This would require an increase of the $^{56}Ni$ mass by a 
factor of $\approx 30$, so this option is excluded. 
A decrease in the rate of change of the \gr\
escape probability would also flatten the light curve.  This
can be achieved by a modification of the structure of 
the inner layers of the ejecta that determine the escape probability.
Either a reduction of the expansion velocity of the inner layers 
or a steepening of the density profile there would give
a slower decline from maximum.  
Both are expected for strongly aspherical explosions because
the energy deposition will not be at the inner edge but distributed
over a larger radial distance.
In principle, late time observations of the width of $^{56} Co$ 
lines and the very late bolometric
light curve could be used to distinguish  the different options.
In light of our approximations, we have chosen to forgo 
the exercise of further fine tuning the models in favor of
future, more self-consistent hydrodynamic and radiative
transfer calculations.

\section{Discussion and Conclusions}
 
We have presented light curves for aspherical models of SNe~Ic.
The size of the asphericity was chosen to match the observed 
polarization in core collapse supernovae with small 
or negligible H-rich envelopes. 
For appropriate axis ratios of about 2,
we find that the apparent luminosity varies from equator to 
pole by roughly one magnitude for both oblate and 
prolate structures. The effect of asphericity may therefore
be as important as the total amount of $^{56}Ni$.
 
Asphericity may be produced by rapid rotation of the progenitor 
in a close binary system or by aspherical explosions. 
An example of the latter would be a greater energy release in 
the polar direction.  Such an asymmetric energy input will produce 
oblate density structures.  More energy release into the
equatorial direction will cause oblate structures. 
 To result in a final asymmetry in the homologous phase 
with an axis ratio $\sim2$ sufficient to
account for the polarization of SN~Ic and the luminosity 
of SN~1998bw requires a factor 
of 6 more kinetic energy per unit solid angle in the 
polar direction compared to the equatorial direction.
Since asphericity tends to be smoothed out during the acceleration 
phase prior to homologous expansion this may imply a much 
larger initial anisotropy in the energetics of the
core collapse process itself. 
 
Because we do not know the mechanisms of core collapse or
the production of asymmetries, it is important to look
for statistical properties of the different configurations.
For oblate geometries, high luminosity should be correlated with 
low polarization and high expansion velocities whereas
the opposite trends are expected for oblate structures. 
 
We have shown that the high apparent luminosity of SN~1998bw 
may be understood within the framework of ``classical" SNIc. 
Even with our current model, SN~1998bw remains 
at the bright end of the scale.  We note that the luminosity of 
SN~1998bw may be uncertain by a factor of $\pm$ 50 percent due to non-Hubble
motion within the cluster and uncertainties in the Hubble 
constant and the reddening. 
For a model with an ejected mass of the C/O envelope
of 2 $M_\odot$, an explosion energy of $2\times10^{51}$ erg, and 
an ejected $^{56}Ni$ mass of $0.2 M_\odot$,
both the bolometric and broad-band light curves 
are rather well reproduced by an oblate 
ellipsoid with an axis ratio of 0.5 which is observed within 
$30\deg$ of the symmetry axis.
This angle for the line of sight is consistent with the low 
(but still significant) polarization observed for SN~1998bw. 
In a Lagrangian frame, the polar expansion velocity is a factor of 2 larger
than the mean velocity. This is also in agreement with the rather 
large expansion velocities seen in SN~1998bw.
Within our framework, prolate geometries can be ruled out both from the
polarization and the expansion velocities.
 
Woosley et al. (1998) have analyzed the possibility of 
$\gamma $-ray bursts in the framework
of spherical models. Even with their explosion energies of more 
than 20 foe they showed that the \grb\ associated with
SN~1998bw/GRB~980425 cannot be explained by the acceleration
of matter to relativistic speeds at shock-breakout. 
In our picture, the specific energy released in the polar region 
is comparable to that in Woosley's models even though the
total energy is much less. 
This asymmetry in the energy distribution, however, corresponds to 
the homologeous phase. 
The initial anisotropy in the energy distribution
associated with shock break-out 
is expected to be significantly higher. 
We note that the high energy densities in a small amount
of mass near the axis could give high entropy in the inner
regions and hence conditions of high entropy that are
conducive to the formation of an r-process.
 
Detailed studies of aspherical explosions, light curves and NLTE-spectral
synthesis will be done in the future to provide a clear separation of 
the different scenarios.
 Note that the time evolution of $P$ depends on the result of the hydrodynamical 
structure, i.e. its geometry,  the variation of the shape of the  'photosphere'
 and the change of the optical depth  (see above and H\"oflich 1995b).
The list of possible improvements to the current models is 
long and includes stellar structures for rapidly
rotation cores, aspherical core collapse
and, in light of the dependence of the post-maximum decline
on the \gr\ escape probability, 
a more realistic distribution of the explosion energy 
in the inner layers.

We have shown that SN~1998bw may be 
understood within the framework of ``classical" core collapse supernovae 
rather than as a ``hypernova". In light of the reasonable fits of 
Iwamoto et al. (1998), however, we cannot rule out the hypernova scenario.  
There may be ways to discriminate the hypernova scenario
from our asymmetric models. Aspherical explosions with kinetic
energy of 1 foe are expected to produce much smaller velocities
for the inner $^{56}Ni$ rich layers that hypernova models with
$\gta$ 20 foe.  The asymmetric models predict rather low
expansion velocities ($v_{exp} \lesssim 7000$ \kms) 
for $^{56}Co$ that could be observed in late time spectra.
Another difference is the polarization.  This is not predicted
in the current hypernova models of Iwamoto et al. (1998)
and Woosely et al. (1998), but they could presumably be 
rendered asymmetric in some way.  With lower ejecta mass
the models we present here predict that the polarization should
start to decline  as the envelope becomes transparent at
$\approx $ day 60 - 80. The more massive hypernova models
would probably be consistent with a later phase of transparency
and hence a later phase of declining polarization.

\subsection*{ACKNOWLEDGMENTS} 

We thank  Ken Nomoto for providing us with the 
broad-band light curve data in 
digital form. This research was supported in part by NSF Grant AST 9528110, 
NASA Grant NAG 5-2888, and a grant from the Texas Advanced Research Program.

\end{document}